\documentstyle[11pt,newpasp,twoside,epsf]{article}
\def\arg#1{{\it#1\/}}

\def\edcomment#1{\iffalse\marginpar{\raggedright\sl#1\/}\else\relax\fi}
\def\einstein{{\it Einstein~}} 
\def\aro{{$\alpha_{\rm ro}$~}} 
\def\aox{{$\alpha_{\rm ox}$~}} 
\def\arx{{$\alpha_{\rm rx}$~}} 
\def\ergs{{erg~cm$^{-2}$s$^{-1}$~}} 
\def\vovm{{$V/V_{\rm m}$}~}
\def\veova{{$V_{\rm e}/V_{\rm a}$}~}  
\def\fxfr{$f_{\rm x}/f_{\rm r}$~}
\def\nupeak{$\nu_{peak}$~}
\def\etal{{\rm et al.}~}
\def\vovmave{{$\langle V/V_{\rm m} \rangle$}~}
\def\vovaave{{$\langle V_{\rm e}/V_{\rm a} \rangle$}~}  
\reversemarginpar

\begin{document}
\title{The Cosmological Evolution of BL Lacertae Objects}
\author{Paolo Giommi, Alberto Pellizzoni}
\affil{Agenzia Spaziale Italiana, Viale Liegi 26, 00198 Roma, Italy}
\author{Matteo Perri}
\affil{BeppoSAX Science Data Center, Via Corcolle 19, 00131 Roma, Italy}
\affil{Dipartimento di Fisica, Universit\`a di Roma ``La Sapienza'', 
       P.le A. Moro 2, 00185 Roma, Italy}
\author{Paolo Padovani}
\affil{Space Telescope Science Institute, 3700 San Martin Dr., Baltimore, 
       MD, 21210, USA}
\affil{Affiliated to the Astrophysics Division Space Science Department, 
       European Space Agency}
\affil{On leave from Dipartimento di Fisica, II Universit\`a di Roma 
       ``Tor Vergata'', Via della Ricerca Scientifica 1, 00133 Roma, Italy}

\begin{abstract}
We review the main results from several radio, X-ray and multi-frequency 
surveys on the topic of cosmological evolution of BL Lacertae objects. 
Updated findings on BL Lac evolution following the recent identification 
of many sources in the ''Sedentary Multi-Frequency survey'' are also 
discussed.
By means of extensive Monte Carlo simulations we test some possible explanations 
for the peculiar cosmological evolution of BL Lacs. We find that a dependence
of the relativistic Doppler factor on radio luminosity (as expected within
the beaming scenario) may induce low values of \vovmave and that both edge effects 
at the low luminosity end of the BL Lacs radio luminosity function, and  
incompleteness at faint optical magnitudes may be the cause of the low \vovmave 
found for extreme HBL sources in X-ray selected samples. 

\end{abstract}

\section{Introduction}
BL Lacertae objects, like quasars, were discovered in the late sixties as unexpected 
extragalactic  counterparts of bright radio sources. Soon after their discovery the 
number of known QSOs (radio loud, and especially the radio quiet ones) grew rapidly 
thanks to the detection of many new objects at optical and radio frequencies. Very 
few BL Lacs could instead be found due to their peculiar characteristics which 
include lack of strong emission lines and absence of 
the ``big blue bump''. The strong optical variability and polarization properties 
that are typical of BL Lacs turned out not to be sufficiently strong features to allow 
the discovery of many new objects of this class at optical 
frequencies. The number of known BL Lacs remained very low for a long time.
When the first generation of X-ray satellites (UHURU, ARIEL V) established that 
emission line 
AGN are strong X-ray emitters also BL Lacs were detected as powerful X-ray sources. 
Shortly afterwards the HEAO1 all sky X-ray surveys discovered many new AGN including a 
few BL Lacs (Piccinotti \etal 1982). 
The small number of BL Lacs compared to QSOs (both at radio and X-ray frequencies),
together with their extreme properties, nicely fitted into the picture where these 
objects are the fraction of relativistically beamed sources that happens to be viewed 
at a small angle with respect to our line of sight (Blandford \& Rees 1978). The total 
number of beamed sources pointing in any direction (the parent population of BL Lacs) 
would then be much larger.

When the \einstein observatory detected X-ray emission form nearly all known AGN and 
discovered many new ones as serendipitous sources, the number of X-ray discovered BL Lacs 
was expected to grow in proportion to that of QSOs. Surprisingly instead (although
some BL Lacs were indeed found as X-ray serendipitous sources establishing that the 
X-ray selection is one of the main methods for the selection of new BL Lacs) the fraction 
of BL Lacs was found to sharply decrease at faint X-ray fluxes. The strong increase of 
space density or luminosity at earlier cosmic epochs (cosmological 
evolution) that was found in the radio, optical and X-ray counts of QSOs was clearly not 
present in the population of BL Lac objects (Stocke \etal 1982; Maccacaro \etal 1983). 
The \einstein Medium Sensitivity Survey (EMSS) convincingly showed that  
X-ray selected BL Lacs display negative cosmological evolution, that is 
either there were less objects or they were less luminous in the past (Morris \etal 
1991; Wolter \etal 1991; Rector \etal 2000). 

In the following paragraphs we review the main results on the cosmological evolution of BL Lacs
as obtained from several surveys in different energy bands. 

\section{BL Lac surveys}
BL Lacs are intrinsically rare objects, therefore it is necessary to select them amongst 
a much larger number of other 'contaminating' sources.
Historically this has been done by means of two main techniques:
\begin{enumerate}
\item BL Lacs are searched for in flux limited samples by means of 
optical spectroscopy of {\it all} the candidates detected above the survey limit. 
This approach can be afforded only if the number of candidates is of the order of 1000 or less.
In the following we will refer to these surveys as 'classical surveys'.

\item The search makes use of a pre-selection achieved through a cross-correlation of sources
detected in surveys carried out in different energy bands, sometimes combined with multi-band 
spectral restrictions.
The candidate BL Lacs can then be identified through optical spectroscopy as in  'classical 
surveys' or through statistical identification techniques. The big advantage of this method is 
that the pre-selection can reduce the number of candidates from several thousands or  
even millions to much smaller samples including only a few  hundreds objects or less.  
\end{enumerate}

\begin{table}[h]
\caption{\arg{Single band flux limited surveys}}
\begin{tabular}{|l|c|c|c|c|}
\hline 
Survey &No. of &Radio flux &X-Ray flux l. & \vovmave \\
 &BL Lacs& limit (mJy) & \ergs & \\
\hline 
Early radio surveys &   & $>$ 1000  & - &  - \\
1Jy complete sample & 34&     1000  & - & $0.60\pm0.05$ \\ 
S4/S5(incomplete)   & 7,11 &  500,250  & - & - \\ 
{\it Einstein} EMSS & 41 &$\approx$ 1&$\approx 2\times 10^{-13}$ &$0.427\pm0.045 $\\
{\it Einstein} Slew Survey & 66 &$\approx$ 1&$\approx 5\times 10^{-12}$ & -\\
RASS HBX BL Lacs & 35 &$\approx$ 2&$ 8\times 10^{-13}$ & $0.41\pm0.05$\\
\hline 
\end{tabular}
\end{table}

\begin{table}[h]
\caption{\arg{Multi-frequency BL Lac surveys }}
\begin{tabular}{|l|c|c|c|c|c|c|}
\hline 
Survey &En. Band &No. of &Vmag &Radio fl &X-Ray flux l.& \vovmave \\
       &limits &objects& limit & limit  & \ergs & \\
       &&& & (mJy)  & & \\
\hline 
RGB  &R/O/X & 127  & 18 & 25 &  $\approx 1\times 10^{-12}$&- \\
REX  &R/O/X & 44  & 20 & 5 &  $4\times 10^{-13}$& $0.52\pm0.04$ \\
DXRBS  &R & 30  & 25 & 50 &  $2\times 10^{-14}$& $0.57\pm0.05$ \\
Sedentary  &R & 162  & 21 & 3.5 &  $\approx 2\times 10^{-12}$&$0.41\pm0.02$ \\
\hline 
\end{tabular}
\end{table}

Table 1 lists the most important classical radio and X-ray surveys together with 
the main parameters and results on the cosmological evolution of BL Lacs. 
These surveys select different types of BL Lacs (High energy 
peaked or HBL, Low energy peaked or LBL, and intermediate objects IBL, see 
Padovani \& Giommi 1995) depending on the energy band and the flux limits.
Given the very limited number of objects in each survey it is not possible 
to study the cosmological
evolution comparing luminosity functions built in different redshift shells. 
The \vovmave estimator, expected to be 0.5 for a non evolving population of 
objects (Schmidt 1968), however provides a suitable tool to study cosmological 
evolution in BL Lacs. Table 1 shows that
\vovmave indicates positive evolution (i.e.  \vovmave $> 0.5$) for bright radio sources
and negative evolution for X-ray selected samples (where radio fluxes are rather low). 
In all cases the statistical significance is about 2 sigma. In the case 
of the RASS HBX sample \vovmave decreases and the significance improves if only the 
extreme HBL objects are considered (Bade \etal 1998). This effect is also present 
in the EMSS sample (Rector \etal 2000). In both cases less extreme objects are 
consistent with no cosmological evolution.

Table 2 lists a number of BL Lac samples selected by means of surveys with flux limits 
in more than one energy band, or where a multi-frequency preselection has been 
applied to define a sample that is flux limited in one band. 
Again the different flux limits tend to select different types 
of BL Lacs. The \vovmave results range from consistency with  no evolution to strong 
negative evolution for extreme HBL (\arx $ < 0.56 $) (Giommi, Menna \& Padovani 1999). 
In the latter survey \vovmave is significantly below 0.5 only at radio fluxes below 10 mJy. 

{\it Strong negative evolution seems to be confined to extreme HBL sources, probably at low
radio fluxes.} 

First results on the REX survey (Caccianiga \etal 2000) however do not seem to confirm 
this trend, although the radio flux limit (5 mJy) in this case is somewhat higher than that 
of the other X-ray surveys. 

The sedentary survey is the most peculiar among those listed above since it does not
rely on direct optical spectroscopy identification of the candidates. In the next paragraph we 
briefly summarize the main characteristics of this survey and we give an update on the status 
of the identification process.

\section{The Multi-frequency Sedentary Survey: an update}

The Multi-frequency Sedentary Survey (Giommi, Menna \& Padovani 1999) 
makes use of a very efficient multi-frequency selection technique to define 
a large (radio flux limited) sample of extreme (\arx $< 0.56 $) High Energy Peaked BL Lacs 
(HBL, Padovani \& Giommi 1995). In the original paper the sample included 
155 objects 58 of which were known BL Lacs and the remaining 97 were still 
unidentified (but with $\approx 85\%$ chance of being BL Lacs). 
Using this first version of the sedentary survey Giommi \etal (1999a) found evidence 
for strong negative evolution at fluxes lower than 10-20 mJy. 
In order to use the best possible estimate of the cosmological 
evolution of BL Lacs from this survey we have updated it by adding 
about 40 new identifications obtained partly through our own optical identification
program and (mostly) from the recent publication of the results of massive 
identification campaigns of X-ray sources discovered in the Rosat All Sky Survey
(Bauer \etal 2000; Schwope \etal 2000). 
As expected a very large fraction of the new identifications are indeed BL Lacs. 
To give an example, of the 27 candidates in the sedentary survey that are in
the list of Schwope \etal (2000) 24 are BL Lacs, 1 is a cluster of galaxies (or 
a BL Lac in a cluster) and only 2 are emission line AGN. 

The updated sample now includes 162 objects resulting from the inclusion of a few 
extra sources that have been identified as BL Lacs just below the \aro limit that was
used to avoid contamination with Seyfert Galaxies (see Giommi \etal 1999a) and  
the removal of the few candidates identified with emission line AGN. 
Of the 108 identified sources (67\% of the sample) 104 are BL Lacs and 4 have 
been classified in the literature as clusters of galaxies (although these could 
well be BL Lacs in a cluster). The remaining 54 candidates are still 
unidentified but, given the very successful identification rate obtained
so far, we are confident that these sources are in very large percentage (85\%-90\%) 
BL Lacs.

The application of the \veova test, a variant of \vovm for surveys with
more than one flux limit (see Avni \& Bachall 1980), gives the results reported 
in table 3.

\begin{table}
\caption{\arg{The Multi-frequency Sedentary Survey: \vovaave results }}
\begin{center}
\begin{tabular}{|l|c|c|}
\hline 
Radio flux limit &\vovaave&No. of objects \\
  (mJy) &  & in sample\\
\hline 
 3.5  & $0.41\pm0.02$ & 162 \\
 5.0  & $0.43\pm0.02$ & 140 \\
 10.0  & $0.45\pm0.03$ & 87 \\
 20.0  & $0.48\pm0.04$ & 45 \\
\hline 
\end{tabular}
\end{center}
\end{table}

The strong indication for negative cosmological evolution (\vovaave $< 0.5 $) at 
low radio fluxes found by Giommi \etal 1999a is confirmed.  
The updated radio logN-logS (not shown here for reasons of space) is also 
very similar to that derived with the original sample of 155 objects. The 
source counts fall sharply below the `Euclidean' slope at low fluxes consistently 
with negative cosmological evolution. 

\section{Possible interpretations}

The most direct explanation of the low \vovmave values of BL Lacs calls for a 
cosmological evolution of these objects that is considerably less than that 
of emission line AGN which have been found to be either more numerous or brighter
in the past than now. The space density of BL Lacs (or their luminosity) could
instead be approximately constant or even decrease at high redshifts. 
A model that predicts a lower amount of cosmological evolution in BL Lacs 
compared to QSO's as a result of changes in the values of fundamental parameters in 
the fueling mechanism of the central engine in AGN is described by 
D'Elia \& Cavaliere (2000).

In the following we explore a different hypothesis, namely that the low values 
of \vovmave might be caused (at least in part) by geometrical and relativistic 
effects expected in the beaming scenario. These effects influence 
the position of the synchrotron peak in the broad-band spectral energy 
distribution of blazars, and hence the composition of flux limited samples 
selected in energy bands above the position of \nupeak.

\begin{figure}[h]
\epsfysize=11.0cm
\hspace{1.5cm}\epsfbox{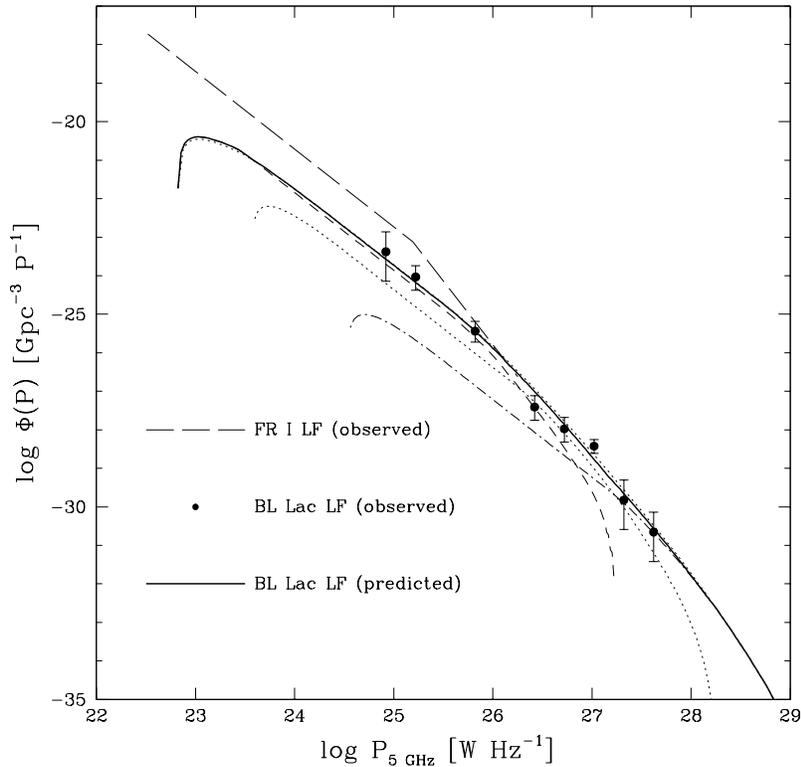}
\caption[h]{The observed and predicted luminosity function of BL Lacs based on the 
beaming model described in Urry \& Padovani (1995). The contribution to the full 
luminosity function (solid line) coming from different beaming amplifications are 
shown as short-dashed ($5<\delta<10$), dotted ($10<\delta<25$) and dot-dashed thin 
lines ($25<\delta<50$).} 
\end{figure}

It is commonly accepted that the broad band electromagnetic spectrum 
of BL Lacs is dominated  by synchrotron emission, up to a peak frequency 
\nupeak (in a $\nu~vs~L(\nu)$ representation) followed at higher energies 
by Inverse Compton radiation produced in a relativistically moving plasma 
seen at a small angle with respect to the line of sight (e.g. Urry and 
Padovani 1995).     
Within this scenario the position of the synchrotron peak is determined by 
the strength of the magnetic field ($B$), the maximum energy of the electron 
spectrum ($\gamma_{max}$) and by the beaming Doppler factor ($\delta$) which 
depends on the bulk motion and on the viewing angle: 

\begin{equation}
\nu_{peak} \propto B~\gamma_{max}^{2}~ \delta
\end{equation}  

\noindent 

The observed range of \nupeak is extremely wide 
spanning from $10^{12}-10^{14} Hz$ for the large majority of BL Lacs selected in radio 
surveys, to $10^{17}-10^{18} Hz$ (and possibly higher) for X-ray discovered 
HBL sources (Pian \etal 1998; Giommi \etal 1999b). Since radio selection 
is not expected to influence the position of \nupeak objects showing low frequency 
peaks are thought to be much more common than HBLs (Padovani \& Giommi 1995). 
Emission line Blazars (also known as Flat Spectrum Radio Quasars FSRQ) also show 
a variety of  \nupeak values but without reaching very high frequencies as 
BL Lacs (Padovani 2000; Perlman \etal 2000).

BL Lacs (like all Blazars) are thought to be the fraction of intrinsically much fainter 
sources that happen to have the beaming axis oriented close to our line of sight. In this 
framework the luminosity function of BL Lacs can be estimated starting from that of 
their parent population (Urry \& Padovani 1995).
Figure 1 shows the observed radio luminosity function (LF) together 
with that obtained by ''beaming'' the LF of FRI radio galaxies, the assumed parent 
population of BL Lacs.  
In the beaming scenario the high luminosity end of the beamed LF is built  
combining the highest luminosities in the parent population with the 
highest Doppler amplification factors ($\delta$). Equation (1) then
implies that, on average, brighter objects will have higher synchrotron peak 
frequencies (\nupeak). At the other 
end of the LF the opposite occurs, namely the low luminosity beamed 
objects will be characterized by lowest $\delta$ factors, and 
consequently by lower \nupeak frequencies. The 
contributions to the beamed LF of low, intermediate and high Doppler 
factors are shown in figure 1 as short-dashed, dotted and dashed-dotted lines 
respectively. 

Figure 2 shows the SEDs of two sources at the bright and 
faint end of the radio luminosity function. Since the $\delta$ factor
is on average fairly large in high luminosity objects, eq (1) implies that
also \nupeak is on average located at higher energies in bright sources. 
The brightness in the X-ray band (marked by the dashed lines) compared to radio 
intensity (\fxfr) is then larger in high radio luminosity objects than in  
low power radio sources. This difference favors the inclusion of powerful  
BL Lacs in surveys of extreme HBLs where the \fxfr ratio is by definition
very high. Near the flux limit 
of a deep radio survey (where an increasing fraction of the BL Lacs comes
from the low luminosity part of the LF) the number of HBLs does not 
grow as fast as the rest of the population because \nupeak gets
smaller and smaller. This effect could flatten the logN-logS of extreme HBLs 
and bias their \vovmave towards low values.

\begin{figure}
\epsfysize=8.0cm
\hspace{1.5cm}\epsfbox{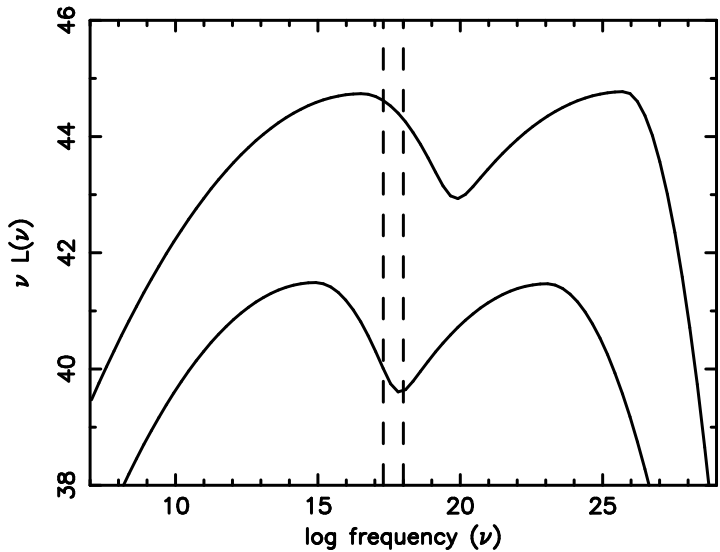}
\caption[h]{Spectral Energy Distributions of BL Lac objects with different \nupeak 
due to different $\delta$ factors at the low and high power ends of the radio
luminosity function. The dashed lines indicate the X-ray band.  
Because of the different \nupeak energies the \fxfr is much higher in 
the high radio luminosity object.}
\end{figure}

\section{Monte Carlo Simulations of BL Lac Surveys}

In order to study in a quantitative way the effects of the
$\delta$ - radio luminosity dependence described in the previous 
section we have carried out extensive simulations of surveys with flux limits 
at radio, optical and X-ray frequencies. In particular we have reproduced 
the major surveys listed in table 1 and 2, and we have applied to them 
the same \vovm tests that have been applied to real data. 

\begin{figure}[h]
\epsfysize=8.0cm
\hspace{1.5cm}\epsfbox{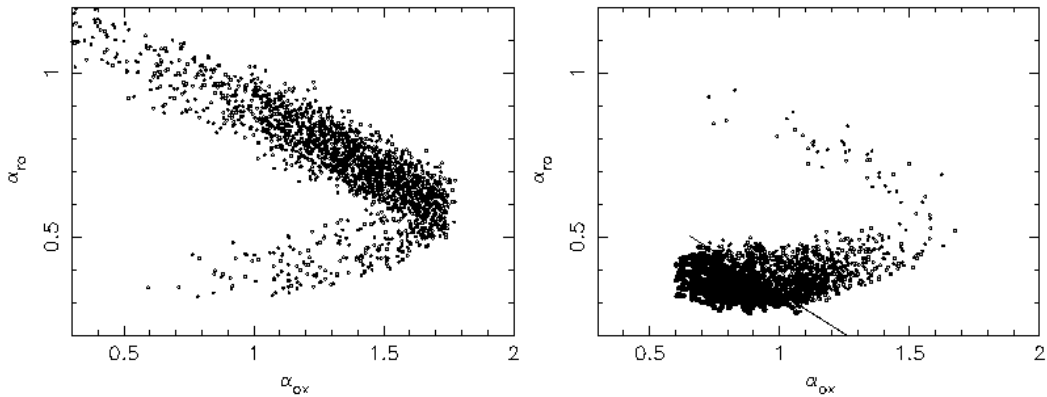}
\caption{Left: Simulation of a radio flux limited survey of BL Lac objects; 
Right: Simulation of BL Lacs in the RASS survey. Filled circles to the left of 
the diagonal line are extreme HBLs as in the multi-frequency sedentary survey.}
\end{figure}

Our procedure can be summarized as follows. 
We start by simulating a redshift value assuming a standard Friedman cosmological model
with $q_o=0$ and $H_0=50~km~s^{-1}~Mpc^{-1}$; then we assign a luminosity to our simulated
source by drawing it from the radio luminosity function of BL Lacs shown in figure 1,
which was obtained by beaming the luminosity function of FRI galaxies (Urry \& Padovani 1995). 
No cosmological evolution is applied, i.e. the evolution parameter is set to 0. 
The source intensities at other frequencies (optical and X-ray in this case, but 
any other energy band can be simulated) are calculated through a Synchrotron Self 
Compton (SSC) model that produces Spectral Energy Distributions similar to that shown in 
figure 2. The simulated source is then accepted if its radio, optical and X-ray 
fluxes are all above the limits chosen for the simulation. 
In calculating the broad band spectrum from the SSC model the value of the Doppler 
factor ($\delta$) (which determines the position of \nupeak in the SED, see 
figure 2) was correlated to the radio luminosity function so that high luminosity sources  
have higher probability to have a large $\delta$ as predicted by the beaming 
model (figure 1). All other SSC parameters (maximum Lorenz factor $\gamma$, 
magnetic field ($B$), the electron spectral slope etc.) were chosen so that 
the prediction of the model are consistent with the observed BL Lac SEDs available 
in the literature. 

\subsection{Simulations results}

Figure 3 shows the \aox - \aro plane of a simulated radio flux limited
survey with $f_r > 1~mJy$ (left panel) and of an X-ray survey limited by 
$f_x > 1.0  \times 10^{-12}$ \ergs and  by radio  flux $f_r > 2.5~mJy$ 
(right panel).

\begin{figure}[h]
\epsfysize=8.0cm
\hspace{1.5cm}\epsfbox{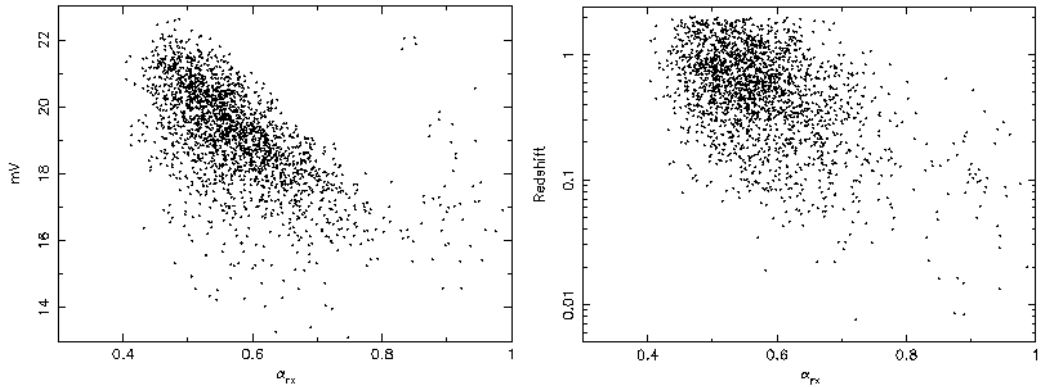}
\caption{The optical magnitudes (left panel) and redshift (right panel) of all 
sources in a simulated X-ray flux limited survey are plotted against the radio 
to X-ray spectral index \arx.}
\end{figure}

This last simulation reproduces the subset of BL Lacs in the Rosat all sky 
survey (RASS) that have radio flux high enough to be included in 
the NVSS survey (Condon \etal 1998).  
Thick circles to the left of the diagonal line represent those objects 
with \arx $< 0.56$ and $f_r > 3.5~mJy$ that 
make a radio flux limited sample of extreme HBLs as in the `sedentary
survey'. 
A sample of 2000 sources was produced in each simulation run. 
We have then analyzed the \vovm distribution in all simulated surveys and
compared the results with those of the `sedentary' and EMSS surveys. 

The results of the \vovm analysis applied to the `sedentary survey'
give \vovmave = $0.45 \pm 0.01$ for the case of a radio flux limit of 3.5 mJy and
\vovmave = 0.50 for a 20 mJy flux limit. This increase of the \vovmave with 
radio flux reproduces very well the observational results reported in
table 3 demonstrating that a dependence of the beaming Doppler factor 
$\delta$ on radio luminosity can induce a bias in the estimation of \vovmave
at low fluxes. 
However we note that these results can be obtained only within a narrow 
range of the parameters that determines how the Doppler factor $\delta$ depends 
on radio luminosity.
This mechanism works in BL Lacs and not in emission line blazars because the 
distribution of synchrotron peak energies is different in the two classes, with 
\nupeak reaching X-ray and possibly higher frequencies in BL Lacs (Giommi \etal 1999b)
and at most the optical/UV band in other blazars (Padovani 2000). 
\begin{table}[h]
\caption{\arg{Simulated X-ray flux limited survey: \vovmave results }}
\begin{center}
\begin{tabular}{|l|c|c|}
\hline 
 \arx $~$limits &\vovmave&No. of objects \\
  & & in sample\\
\hline 
All values& $0.49(0.47)$ & 2000(1900) \\
 $< 0.55$ & $0.47(0.43)$ & 852(749) \\
 $< 0.50$ &  $0.44(0.37)$ & 348(279) \\
 $< 0.48$ &  $0.42(0.35)$ & 204(160) \\
\hline 
\end{tabular}
\end{center}
\end{table}

We have next considered the case of a `classical' X-ray flux limited survey like 
the EMSS. To keep things simple we have chosen a single X-ray flux limit equal to 
$5\times 10^{-13}$ \ergs rather than a sky coverage with multiple flux limits. 
We do not expect significant differences between the two cases.
The \vovmave that we obtained for the total sample is 0.49, only slightly lower 
than the expected value of 0.5 in case of no dependence between $\delta $ and 
luminosity. 
We have then verified whether our simulated survey shows the strong dependency 
of \vovmave on the \fxfr ratio (or, equivalently on \arx) found by Rector \etal 2000. 
Table 4 shows the \vovmave values for increasing values of \arx. A clear
dependency of the \vovmave  on \arx is confirmed. This dependence, however is 
still present even if $\delta $ is set to be constant for all simulated sources. The decrease
of \vovmave at high \fxfr values in our simulation is due to the fact that high \fxfr sources 
close to the X-ray survey limit must be weak radio sources near the low luminosity end 
of the radio luminosity function. At even higher \fxfr values the radio luminosity becomes  
lower than the low luminosity limit of our LF and the number of sources drops. 
Another, possibly more important, reason for the observed \vovm dependency on \fxfr could be 
incompleteness of the sample at very faint optical magnitudes. 
In a X-ray flux limited sample high \fxfr (low \arx) sources tend to have faint optical 
magnitudes as shown in the left part of figure 4 where the \arx is plotted against mV. 
A small incompleteness, possibly arising from the misclassification of optically very faint 
BL Lacs with cluster of galaxies or radio galaxies, introduces a spurious correlation 
between \fxfr and \vovmave. We have simulated this incompleteness by removing 
5\% of the sources at the optically faint end of the simulated survey. The results are reported 
in parenthesis in table 4 where we can see that the decrease of \vovmave 
with \fxfr becomes considerably higher than in the complete sample. 
The right plot in figure 4 shows a clear correlation between redshift 
and \arx showing that extreme HBL sources are expected to be on average at higher 
redshift that less extreme BL Lacs as found by Bade \etal (1998).

\section{Conclusions}

Various surveys at radio and X-ray frequencies show that the amount of cosmological 
evolution in BL Lacs is significantly lower than that found in other types of AGN. 
The results however vary from no evolution (or slightly positive 
evolution) at high flux values to strong negative evolution at faint fluxes, 
especially for extreme objects with very high \fxfr. 

By means of extensive Monte Carlo simulations we have explored various 
causes that might contribute to explain these findings. 
The following results were obtained.
 
\begin{itemize}
\item A dependency of the beaming Doppler factor $\delta$ on radio luminosity (expected in the 
relativistic beaming scenario) can at least partially
explain the low \vovmave values found in some X-ray and radio flux limited
samples.
\item Edge effects due to the flattening at the low luminosity end of the radio 
luminosity function of BL Lacs can be one of the reasons for the correlation of 
\vovmave and X-ray loudness (\fxfr).
\item Even small amounts of incompleteness at very faint optical fluxes induce
a correlation between \vovmave and \fxfr similar to that seen in the EMSS and 
RASS HBX Surveys.   
\end{itemize}


\end{document}